\documentstyle[12pt]{article}
\newcommand{\be}{\begin{equation}}
\newcommand{\ee}{\end{equation}}
\newcommand{\bea}{\begin{eqnarray}}
\newcommand{\eea}{\end{eqnarray}}
\newcommand{\nn}{\nonumber}

\newcommand{\bfpi}{\mbox{\boldmath $\pi$}}

\begin{document}

\def\theequation{\arabic{section}.\arabic{equation}}
\def\thesection{\Roman{section}}
\def\thesubsection{\Alph{subsection}}

\setcounter{section}0
\setcounter{subsection}0
\setcounter{equation}0

\author{\sf S. A. Gogilidze, \footnotemark[1] ~
A. M. Khvedelidze, \footnotemark[2]~  V. N. Pervushin \\
{\it Bogoliubov Laboratory of Theoretical Physics,}\\
{\it Joint Institute for Nuclear Research, 141980, Dubna,Russia }}


\title{\bf ON ABELIANIZATION OF FIRST CLASS CONSTRAINTS  }

\date{}
\maketitle
\sf
\footnotetext[1]{${}^{*}$
Permanent address: Tbilisi State University,
380086, Tbilisi, Georgia.}

\footnotetext[2]{${}^\dagger$
Permanent address: Tbilisi Mathematical Institute,
380093, Tbilisi, Georgia\\
\hspace*{0.85cm}Electronic address: khved@theor.jinrc.dubna.su}
\bigskip
\bigskip

\begin{quote}
The systematic method for  the conversion of first class constraints  to
the equivalent set of Abelian
one based on  the Dirac equivalence transformation is developed.
The representation for the corresponding matrix performing this
transformation is proposed. This representation allows one to lead the
problem of abelianization to the solution
of a certain system of  first order {\it linear } differential
equations for  matrix elements .
\end{quote}

\vspace{1.5cm}


\centerline{JINR Preprint E2-95-131}

\newpage

\vspace*{2cm}

\section{\sc Introduction}

It is the purpose of this note to describe a practical method for
the  conversion of non - Abelian constraints into the
Abelian form in the theories with  first class constraints.
For explanation of
the  practical importance of this  procedure
let us briefly recall the general principles of the
description of the standard Dirac theory of Hamiltonian  systems with
constraints ~\cite{DiracL} - \cite{Gitman}.

For the sake of simplicity as usually we will discuss the main ideas
using a mechanical system,  i.e. systems  with a finite number of degrees of
freedom, with having in mind that the direct extension of the results
to a field theory  in general is possible only in the local sense.

Suppose that the  system with \( 2n \) - dimensional
phase space \(\Gamma \) acquires the  following set of irreducible
 first class constraints
\( \varphi_\alpha (p,q), \quad ( \alpha = 1,2, \dots, m ) \)
\bea                \label{eq:constr}
\varphi_\alpha (p,q) \,& = & \,0, \nn\\
\{ \varphi_\alpha (p,q), \varphi_\beta  (p,q)\}\, & = &\,
f_{ \alpha\beta \gamma}  (p,q) \varphi_\gamma  (p,q)
\eea
This means that the dynamics of our system is constrained on
the certain submanifold
\footnotemark  of the total phase space or, in another words,
not all canonical  coordinates  are  responsible for the dynamics .
\footnotetext{
{In the next we will symbolize by notation \({\Gamma}_{c} \)
this \( 2n - m \) --- dimensional submanifold
of  \( {\Gamma} \), \( {\Gamma}_{c} \subset {\Gamma}\) }}
The generalized  Hamiltonian dynamics is described by the extended
Hamiltonian which is a  sum
of  canonical Hamiltonian  \( H_C (p,q)\)  and  a
linear combination of  constraints with arbitrary
multipliers \( u_\alpha (t) \)
\be
H_E (p,q) = H_C (p,q) + u_\alpha (t) \varphi_\alpha (p, q) \,
\ee
The arbitrariness of the functions  \( u_\alpha \)  reflects the
presence in the theory of coordinates whose  dynamics is
governed in an arbitrary way .
According to the principle of gauge invariance for  physical
quantities, these coordinates do not affect  them and thus
can be treated as ignorable.  The main problem that arise is
identification of these coordinates .
If there are in the theory only  Abelian constraints
\be \label{eq:ab1constr}
\{ \varphi_\alpha (p,q),\varphi_\beta  (p,q)\}\,  = \, 0,
\ee
one can find these ignorable coordinates as follows.
In this case, it is always possible \cite{Levi-Civita} -
\cite{Shanmugadhasan} to  define a canonical transformation to a new
set of canonical coordinates
\bea \label{eq:cantr}
q_i & \mapsto & Q_i = Q_i \left ( q_i , p_i \right ),\nn\\
p_i & \mapsto & P_i = P_i \left ( q_i , p_i \right ),
\eea
so  that  \( m \) of the new  \( P  \) 's
 ( \(\overline{P}_1, \dots ,\overline{P}_m \) )
become equal to the abelian constraints (\ref{eq:ab1constr})
\be
\overline{P}_\alpha = \varphi _\alpha \left ( q_i , p_i \right )
\ee
In the new coordinates we have the following system of
canonical equations
\bea
 \dot{Q}^{\ast} & = & \{{Q}^{\ast}, H_{Ph} \}, \nn\\
 \dot{P}^{\ast} & = & \{{P}^{\ast}, H_{Ph} \}, \nn\\
 \dot{\overline{P}} & = & 0, \nn\\
 \dot{\overline{Q}} & = & u(t) ,
\eea
with arbitrary functions \( u(t) \) and with
the physical  Hamiltonian
\[
 H_{Ph}\equiv H_C(p,q)\, \Bigl\vert_{ \varphi_\alpha = 0}
 \equiv H_C (P^{\ast},Q^{\ast}) \Bigl\vert_{ \overline{P}_\alpha = 0}
\]
The physical Hamiltonian \( H_{Ph} \) depends only on the remaining
\( n-m \) pairs of the new canonical coordinates \(
( {Q}^{\ast}_1, {P}^{\ast}_1
 \dots ,
{Q}^{\ast}_{n-m}, {P}^{\ast}_{n-m} )\)
wich are a gauge invariant physical variables.
This means that the coordinates \( \overline{Q}_\alpha \)
conjugated to the momenta \({\overline{P}}\)
are ignorable coordinates and the  canonical system  admits
explicit separation of phase space into sectors:  physical and
nonphysical one
\be \label{eq:str}
2n \left \{ \left (
\begin{array}{c}
q_1\\
p _1\\
\vdots\\
q_n\\
p _n\\
\end{array}
\right )  \right.
\quad    \mapsto  \quad
\begin{array}[c]{r}
{ 2(n-m)\,\, \left\{ \left(
\begin{array}{c}
Q^\ast \\
P^\ast
\end{array}
\right ) \right. }\\
\vphantom{\vdots} \\
{ 2m\, \left\{ \left(
\begin{array}{c}
\overline{Q}\\
\overline{P}
\end{array}
\right ) \right. }
\end{array}
\quad
\begin{array}[@]{c}
{ \begin{array}{c}
 Physical \\
sector
\end{array}
} \\
\vphantom{\vdots} \\
{ \begin{array}{c}
Nonphysical\\
sector
\end{array}}
\end{array}
\ee
The straight generalization of this method to the non - Abelian case is
unattainable;
identification momenta with constraints is forbidden due to the non - Abelian
character of constraints (\ref{eq:constr}).
However, there is a general proof of a possibility  of
a local replacement of  the constraints (\ref{eq:constr})
by the equivalent set of constraints  forming  an abelian
algebra ~\cite{Sunder}, \cite{Gitman}, \cite{Shanmugadhasan},
\cite{Maskawa}, \cite{Henn}.
This general
observation --- {\it Abelianization statement  \/} reads ---
\begin{quote}
{\it
For a  given set of \(m \) first class constraints  it is always possible to
chose locally  \(m \)  new equivalent constraints
 \be
 \varphi_\alpha
(p,q) \, =  \,0  \Longleftrightarrow \Phi_ \alpha (p,q ) \,=\, 0 \;
\ee
that define the same constraint  surface \( \Gamma_C \) so  the Poisson
brackets between the new constraints strongly vanishes, i.e. }
\end{quote}
\be
\{ \Phi_\alpha (p,q),  \Phi_\beta  (p,q)\}\,  = \, 0.
\ee
Thus, one  can deal with this equivalent set of Abelian constraints to
construct the
reduced phase space, the space of physical degrees of freedom.  To
reveal ignorable coordinates, we need an explicit form of the new Abelian
constraints \(\Phi_\alpha (p,q) \).
We would like to emphasize that in all proofs of the abelianization it
has been assumed that the constraints form the functional group
\cite{Sunder}, \cite{Henn}.
In the present paper, we shall
point out
two alternative schemes of realization of the abelianization procedure:  via
constraints resolution  and via  ``generalized'' canonical
transformation for general non - Abelian constraints of type
(\ref{eq:constr}).
The generalized canonical transformations ~\cite{Bergman}
are those preserving the form of all constraints
of the  theory as well as the canonical form of the equations of motion .
It will be shown that in a constructive fashion it is possible
to convert constraints into the
Abelian form with the  help of the Dirac  equivalence linear
transformation
\be \label{eq:sufcon}
\framebox[90mm]{ \raisebox{0ex}[4ex][3ex]{
$  \Phi_ \alpha (p,q ) = {\cal D}_{\alpha \gamma}\varphi_\gamma(p,q)
$}}
\ee
with the nonsingular  matrix \({\cal D} \)
\[
det||{ \cal D}_{\alpha \gamma} ||\, \Bigl\vert_{{constraints}} \;\;\not = 0
\]
The main point of our result is that this
matrix \({ \cal D}\)
can be determined by {\it linear } first order differential equations.

The  remaining  part of this note is the proof of this statement and
the   application  to the special  example:  non -  Abelian
Christ and Lee model ~\cite{Christ}

\bigskip
\section{\sc Abelianization : alternative schemes}

\subsection{\sf{Abelianization via constraint resolution}}
\bigskip

The direct way of abelianization of constraints is  as follows
\cite{Henn}, \cite{Newman}.
Under the assumption that \(\varphi_\alpha (p,q) \)  are \( m \)
independent functions one can resolve the constraints
(\ref{eq:constr}) for \( m \) of \( p \)'s
\be
p_\alpha = F_\alpha (\underline{p}, q)
\ee
where  \(\underline{p} \) denotes the remaining  \( p\)'s.
Let us pass  to a new equivalent to the \(\varphi_\alpha (p,q) \)
constraints
\be \label{eq:modcon} \Phi_ \alpha
(p,q ) = p_\alpha  - F_\alpha (\underline{p}, q)
\ee
By  the explicit computing one can convince  that
the Poisson brackets of the new constraints
 \[\{\Phi_ \alpha (p, q) ,\Phi_ \beta (p,q ) \}
\]
are independent of \(p_\alpha\),
but as it is clear that  they are again the
first class the unique possibility is
that  their Poisson brackets with each other
must vanish identically .  Thus, after  transformations  to the  new
constraints
\(\Phi_ \alpha (p, q)  \) we can realize the above - mentioned canonical
transformation (\ref{eq:cantr}) such  that  \( m \) of the new  \( P  \) 's
become equal to the modified constraints  \( \Phi_\alpha \) (\ref{eq:modcon})
\be
\overline{P}_\alpha = \Phi_\alpha \left ( q_i , p_i \right )
\ee
with
the corresponding conjugate ignorable coordinates \( \overline{Q}_\alpha \).

\bigskip

\subsection{{\sf Abelianization of constraints via Dirac` s transformation }}
\bigskip

In this section, it will be demonstrated that due to the freedom in
the representation  of constraint surface \( \Gamma_c \)
\bea                \label{eq:constr2}
   \varphi_\alpha (p,q) \,& = & \,0 , \nn\\
 \{ \varphi_\alpha (p,q),  \varphi_\beta  (p,q)\}\,&  = &\,
f_{ \alpha\beta \gamma}  (p,q) \varphi_\gamma  (p,q).
\eea
one can always pass with the help of Dirac 's transformation
>from  these  first class non - Abelian  constraints
to the equivalent ones
\be                \label{eq:Diractran}
\Phi_\alpha (p,q) \, = {\cal D}_{\alpha \beta} (p, g) \varphi_\beta  (p,q),
\ee
so that the new constraints will be  Abelian
\bea                \label{eq:abconstr}
    \{ \Phi_\alpha (p,q),  \Phi_\beta  (p,q)\}\, &  = &\, 0.
\eea
According to (\ref{eq:abconstr}) the matrix \({\cal D}_{\alpha \beta}  \)
must satisfy the set of the {\it nonlinear} differential equations
\bea                \label{eq:non}
    \{ {\cal D}_{\alpha \gamma} (p, g)\varphi_\gamma (p,q)  ,
{\cal D}_{\beta \sigma} (p, g)\varphi_\sigma  (p,q)
\}\, &  = &\, 0.
\eea
Such a formulation of the abelianization  statement  means a possibility to
find a particular solution for  this very complicated system of
{\it nonlinear} differential equations.
Beyond the question  eq. (\ref{eq:non}) in this form does not contain
any practical value but, as it will be shown here, there is a particular
solution to this equation which can be  represented as
\be                \label{eq:mattran}
{\cal D} \, = \underbrace{{\cal D}^1 (p, q) \cdots {\cal D}^m (p, q)}_{m}
 \ee
where each matrix \({\cal D}^k \) has a form of the product of
\(k \)'s  \( m \times m \)  matrices
\be                \label{eq:matrix}
{\cal D}^k \, = {\cal R}^{a_k +k} (p, q)\prod_{i=k-1}^{0}{\cal S}^{a_k +i}
(p, q)
\ee
\((a_k \equiv \frac{k(k+1)}{2})\)
\bigskip
\bea \label{eq:r}
\begin{array}{lr}
{}~~~~~~~~~~~~~~\overbrace{~~~~~~~~~~~~~~~}^{k}\overbrace{~~~~~~~~~~~~~~~~~~}^{m-k} \\
{\cal R}^{a_k+k} = \left (
\begin{array}{cc}
{\fbox{\raisebox{0.mm}[10mm][5mm]{ \quad {\mbox{\huge I}} \quad}}}  &
\mbox{\huge 0}      \\
\mbox{\huge 0}     &
\fbox{\raisebox{0.mm}[10mm][5mm]{ \quad ${\mbox{\huge B}}^{a_k+k}$ \quad}}
\end{array}
\right )
\end{array}
\eea
\bea \label{eq:s}
\begin{array}{lr}
{}~~~~~~~~~~~~~\overbrace{~~~~~~~~~~~~~~~~~~~~~~~~~~~}^{k}
\overbrace{~~~~~~~~~~~~~~~~~~~~~~~~~}^{m-k} \\
{\cal S}^{a_k+i} =
\left (
\begin{array}{ccccc|ccccc}
1       &   0      &  \cdots &   0      &   0      &    0       &    0
&\cdots  &   0     &    0   \\
0       &   1      & \cdots  &   1      &   0      &    0       &    0
&\cdots  &   0     &    0   \\
\vdots  &  \vdots  & \ddots  &  \vdots  &  \vdots  &  \vdots & \vdots   &\ddots
  &  \vdots &  \vdots  \\
0       &  0       &  \cdots &   1      &   0      &    0      &    0
&\cdots   &   0     &    0    \\
0       &  0       &  \cdots &   0      &   1      &    0     &    0
&\cdots   &   0     &    0    \\
\hline
0       &  \cdots  &  C_{k+1}^{a_k+i}  &   \cdots  &   0      &      1     &
0     & \cdots  &   0     &   0    \\
0       &  \cdots  &  C_{k+2}^{a_k+i}  &   \cdots  &   0      &      0     &
1     & \cdots  &   0     &   0    \\
\vdots  &  \vdots  & \vdots          &   \vdots  &  \vdots  &  \vdots & \vdots
& \ddots  & \vdots  &  \vdots \\
0       &  \vdots  & C_{m-1}^{a_k+i} &   \cdots  &   0      &     0      &  0
   &  \cdots          &    1    &   0   \\
0       &  \cdots  &  C_m^{a_k+i}    &   \cdots  &   0      &     0      &  0
   & ~~~\cdots ~~~ &    0    &   1
\end{array}
\right ) \\
{}~~~~~~~~~~~~\underbrace{~~~~~~~~~~~~~~~}_{k-i} &
\end{array}
\eea
and the corresponding matrix elements satisfy a set of first order {\it
linear } differential equations ( see below (\ref{eq:lin1}) -
(\ref{eq:lin2})).  Just the linear character of these equations alows one to
speak about a practical usage of the proposed method of abelianization.  As it
will
be explained below, the constraints which are obtained as a
result of the action
of \( k\)'s matrices (constraints at the \(a_k + k\) -th step )
\be
\Phi_\alpha^{a_k+k} =\left ({{\cal D}^k \cdot{\cal D}^{k-1}
 \cdots {\cal D}^1}\right )_{\alpha \beta}  \Phi_\beta^{0}
\ee
obey the algebra where \(k\) constraints have  zero Poisson brackets
with any one.
 From the algebraic standpoint  the proposed method of abelianization  is
nothing but an iterative procedure of constructing  ``equivalent''
algebras \({\cal A}^{a_i}\) of constraints \( \Phi_\alpha^{a_i} \).
In   \(a_m \)  steps
the \(m \)  - dimensional non - Abelian algebra is converted
into an equivalent  Abelian one
in such a manner that  at the \(a_k\) - th step
the obtained algebra \( {\cal A}^{a_k} \) possesses a center
with \( k\) elements
\( {\cal Z}_k [ A ] = \left( \Phi_1^{a_k}, \Phi_2^{a_k}, \dots,
\Phi_k^{a_k} \right) \).
\be \label{eq:setalg}
\framebox[140mm]{ \raisebox{0ex}[4ex][3ex]{
$ {\cal A}^{0}\underbrace{\stackrel{{\cal S}^{1}}{\to}{\cal A}^{1}
\stackrel{{\cal R}^{2}}{\to}}_{{\cal D}^1}{\cal A}^{2}
\underbrace{\stackrel{{\cal S}^{3}}{\to}
{\cal A}^{3}\stackrel{{\cal S}^{4}}{\to}{\cal A}^{4}
\stackrel{{\cal R}^{5}}{\to}}_{{\cal D}^2}{\cal A}^{5}
\dots \underbrace{\stackrel{{\cal S}^{a_k}}{\to}
{\cal A}^{a_k}\dots
\stackrel{{\cal R}^{a_k+k }}
{\to}}_{{\cal D}^k}{\cal A}^{a_k+k} \dots
$}}
\ee
The matrix \( {\cal D}^k\) converts the algebra \({\cal A}^{k}\)
into the algebra \({\cal A}^{k+1}\) in which the center contains
one element more than the previous one.

The proof of the validity  of the representation  (\(\ref{eq:matrix} \))
and the equations for the matrices  \({\cal S}\) and \({\cal R} \)
are obtained by induction.
Suppose that
\( \Phi_\alpha^{a_k} \) - are a set of constraints
(obtained as a result of the
action of \(k-1 \) matrices \( {\cal D}^i\) )
with  algebra having the  center
\( {\cal Z}_k [ A ] = \left( \Phi_1^{a_k}, \Phi_2^{a_k}, \dots
\Phi_k^{a_k} \right) \)
than  a matrix
\( {\cal D}^k\)  from (\ref{eq:mattran}) perform the transformation
to the  new constraints
\be
\Phi_\alpha^{a_{k+1}-1} ={{\cal D}^k }_{\alpha \beta } \Phi_\beta^{a_k+1}
\ee
which form  the algebra with  the center
\( {\cal Z}_{k+1} [ A ] = \left( \Phi_1^{a_{k+1}}, \Phi_2^{a_{k+1}},~ \dots ,
{}~\Phi_k^{a_{k+1}}, \right.\\
\left. \Phi_{k+1}^{a_{k+1}} \right) \)
if the matrices  \({\cal S}, {\cal R} \)
are the solutions to the following set of  linear differential
equations:
\bea \label{eq:lin1}
\left.
\begin{array}{rcr}
\{ \Phi_{1}^{a_k+i-1}, S_{\alpha_k}^{a_k+i} \}& =& 0  \\
\vdots~~~~~~~~~~~~~~~~~~\vdots  &~~~ &\vdots \\
\{ \Phi_{k-1}^{a_k+i-1}, S_{\alpha_k}^{a_k+i} \}& =& 0 \\
\end{array}\right\}\;\;\; \Longrightarrow
\{ \Phi^{a_k+i-1}_{{\overline{\alpha}}_k}, S_{\alpha_k}^{a_k+i} \} = 0
\eea

\bea
\{ \Phi_k^{a_k+i-1}, S_{\alpha_k}^{a_k+i} \}& = &
f^{a_k+i-1}_{k \alpha_k \gamma_k}
  S_{\gamma_k}^{a_k+i}  - f^{a_k+i-1}_{k \alpha_k i+1}
\eea

\bea
\left.
\begin{array}{rcr}
\{ \Phi_{1}^{a_k+k-1}, B_{\alpha_k\beta_k}^{a_k+k}  \}& =& 0  \\
\vdots~~~~~~~~~~~~~~~~~~\vdots  &~~~ &\vdots \\
\{ \Phi_{k-1}^{a_k+k-1}, B_{\alpha_k\beta_k}^{a_k+k} \}& =& 0 \\
\end{array}\right\}\;\;\; \Longrightarrow
\{ \Phi^{a_k+k-1}_{{\overline{\alpha}}_k}, B_{\alpha_k\beta_k}^{a_k+k} \} = 0
\eea

\bea \label{eq:lin2}
\{ \Phi_k^{a_k+k-1}, B_{\alpha_k\beta_k}^{a_k+k} \}  =
- f^{a_k+k-1}_{k  \gamma_k \beta_k}B^{a_k+k}_{\alpha_k \gamma_k}
\eea
where
\(\alpha_k = k+1,\dots, m \;,{\overline{\alpha}}_k = 1,2,\dots, k-1 \)
and  \(f^{a_k+i}_{\alpha  \gamma \beta}\) are the
structure functions of constraints algebra \(A^{a_k+i}\)  at the
\({a_k+i}\) -th step.

Let us begin in  a consecutive order the construction of algebras
with the center containing \(1,2, \dots., m \)  elements.
   For determination of the new algebra with one central element (let \(
\varphi_1\)) one can act in  the following way :
\begin{itemize}
\item[a)]
>from the beginning  exclude  \( \varphi_1\) from the left hand side of eq.
(\ref{eq:constr2}); \item[b)] then  realize abelianization with all others
\end{itemize}
To achieve first, we  can perform the transformation with the matrix
 \({\cal S}^1 \)
\[
\Phi^1_\alpha = {\cal S}^1_{\alpha \beta}\varphi_\beta
\]
of type (\ref{eq:s})
\be
{\cal S}^1 =  \left (
\begin{array}{ccccc}
1        &   0      &   0    & \cdots  &   0    \\
C_2      &   1      &   0    & \cdots  &   0    \\
C_3      &   0      &   1    & \cdots  &   0     \\
\vdots   &   \vdots & \vdots & \ddots  & \vdots  \\
C_m      &   0      &   0    &  \cdots  & 1
\end{array}
\right )
\ee
In the expanding form it is
\bea
\Phi^1_1\, & = &\Phi^0_1 =\varphi_1 \nn\\
\Phi^1_{\alpha_1}& =& \varphi_{\alpha_1} +
C^1_{\alpha_1}\varphi_1
\eea
The new constraints algebra remains the algebra of first class
\bea
\{\Phi^1_{1},  \Phi^1_{\alpha_1}  \}&  = &
f^1_{1 \alpha_1 1}  \Phi^1_1 +
f^1_{1 {\alpha_1} \gamma_1}  \Phi^1_{\gamma_1}\nn\\
\{\Phi^1_{\alpha_1},  \Phi^1_{\beta_1}  \}&  = &
f^1_{ {\alpha_1} {\beta_1} {1}}  \Phi^1_{1}
+ f^1_{ \alpha_1 \beta_1 \gamma_1}  \Phi^1_{\gamma_1}
\eea
and the new structure functions \(f^1_{ \alpha \beta \gamma} \)
are determined through  the old one  \(f_{ \alpha \beta \gamma} \) and
the transformation functions \(C^1_{\alpha_1}\) as follows

\bea \label{eq:str1}
f^1_{1 {\alpha_1} 1}& = &f_{1 {\alpha_1} 1} + f_{1 {\alpha_1} \gamma_1}
C^1_{\gamma_1} + \{\Phi^0_{1},  C^1_{\alpha_1}  \}\\
f^1_{\alpha_1 {\beta_1} 1}& = &{{1}\over{2}} \left(\;
f_{{\alpha_1}{\beta_1} 1} -
 f_{{\alpha_1}{\beta_1} \gamma_1}C^1_{\gamma_1}
+ \{ C^1_{\alpha_1}, C^1_{\beta_1}\}\Phi_1^0 \;\right)  - \nn\\
&-& f^1_{1 {\alpha_1} 1} C^1_{\beta_1} +
\{\Phi^0_{\alpha_1}, C^1_{\beta_1} \}
-(\alpha_1 \leftrightarrow \beta_1 ) \\
f^1_{{\alpha_1}\beta_1 \gamma_1}& =& f_{{\alpha_1}\beta_1 \gamma_1}
+ C^1_{\alpha_1}f_{1 {\beta_1} \gamma_1 } -
C^1_{\beta_1}f_{1 {\alpha_1} \gamma_1 } \\
f^1_{1 {\alpha_1} \gamma_1}& =& f_{1 {\alpha_1} \gamma_1}
\eea
Let us now choose the transformation functions \(C^1_{\beta_1}\)
so that the Poisson bracket of first constraints \( \Phi_1^1 \) with
all other modified constraints do not contain it
\bea
\{ \Phi^1_1 (p,q),  \Phi^1_{\alpha_1} (p,q)\}\,&  = &\,
\sum_{\gamma \not= 1} f^1_{ 1 \alpha_1 \gamma}  (p,q) \Phi^1_\gamma  (p,q).
\eea
These \(m-1 \) requirements \( f^1_{1 {\alpha_1} 1}  = 0  \),
according to (\ref{eq:str1}), mean that the transformation function
\( C^1_{\alpha_1}\) must satisfy  the
following set of linear nonhomogeneous differential equations
\bea \label{eq:ng1}
\{\Phi^0_{1},  C^1_{\alpha_1} \} =
- f_{1 {\alpha_1} 1} + f_{1 {\alpha_1} \gamma_1}
C^1_{\gamma_1}
\eea
Note that the problem of existence of solution to
such a set of equations is studied very well
( see e.g. \cite{Kur} )
Suppose, we find some particular solution \( C^1_{\alpha_1}\)for
(\ref{eq:ng1}), then one can determine all structure functions
of the modified algebra according to eq.(\ref{eq:str1}),
\bea \label{eq:str11}
f^1_{1 {\alpha_1} 1} & = & 0 \\
f^1_{\alpha_1 {\beta_1} 1}& = &
f_{{\alpha_1}{\beta_1} 1} -
 f_{{\alpha_1}{\beta_1} \gamma_1}C^1_{\gamma_1}
+ \{ C^1_{\alpha_1}, C^1_{\beta_1}\}\Phi_1^0\nn\\
&+& \{\Phi^0_{\alpha_1}, C^1_{\beta_1} \}
+\{\Phi^0_{\beta_1}, C^1_{\alpha_1} \} \\
f^1_{{\alpha_1}\beta_1 \gamma_1}& =& f_{{\alpha_1}\beta_1 \gamma_1}
+ C^1_{\alpha_1}f_{1 {\beta_1} \gamma_1 } -
C^1_{\beta_1}f_{1 {\alpha_1} \gamma_1 } \\
f^1_{1 {\alpha_1} \gamma_1}& =& f_{1 {\alpha_1} \gamma_1}
\eea

Now let us again keep first constraint unchanged and
 perform  the Dirac transformation
on the remaining  part of constraints
\( \Phi_{\alpha_1},\;\; \alpha_1 = 2,3,..., m \)
\bea
\Phi^2_1 & = & \Phi^1_1 = \Phi^0_1 =\varphi_1 \nn\\
\Phi^2_{\alpha_1} & = & B^2_{\alpha_1 \beta_1}
\Phi^1_{\beta_1}
\eea
with the requirement that the new constraints have zero Poisson
brackets with the first one \(\Phi^1_1 \)
\bea
    \{ \Phi^2_1 ,  \Phi^2_{\alpha_1}\}\, &  = &\, 0.
\eea
One can verify that this requirement means that the transformation
functions \(B_{\alpha_1 \beta_1} \) are the solution to the equation
\bea \label{eq:h2}
\{\Phi^1_{1},  B^2_{\alpha_1\beta_1} \} =
- f_{1 {\gamma_1} \beta_1}B^2_{\alpha_1\gamma_1}
\eea
With the help of a solution of eq. (\ref{eq:h2})  the modified algebra
has the following structure functions
\bea \label{eq:str2}
f^2_{1 {\alpha_1} 1} & = & 0, \\
f^2_{\alpha_1\beta_1 1}&=& B^2_{\alpha_1\delta_1}B^2_{\beta_1 \sigma_1}
f^1_{\delta_1 \sigma_1 1 }, \\
f^2_{\alpha_1 {\beta_1} \gamma_1}& = &
\left[ \{ B^2_{\alpha_1 \delta_1}, B^2_{\beta_1 \sigma_1}\}
\Phi^1_{\sigma_1}
+
\{ B^2_{\alpha_1 \delta_1}, \Phi^1_{\sigma_1}\}
B^2_{\beta_1 \sigma_1}\right. - \nn\\
&-&\left.
\{ B^2_{\beta_1 \delta_1}, \Phi^1_{\sigma_1}\}
B^2_{\alpha_1 \sigma_1}
+
B^2_{\alpha_1\kappa_1} B^2_{\beta_1 \sigma_1}f^1_{\kappa_1\sigma_1\delta_1 }
\right](B^2)^{-1}_{\delta_1 \rho_1}
\eea
Thus, as a result of two transformations
\({\cal D}^{1} = {\cal S}^{1}{\cal R}^{2} \)
we obtain the  modified algebra \( {\cal A}^2 \)
of constraints
\( \Phi_\alpha^2 \) with the central element   \( \Phi^2_1 \)
\bea \label{eq:alg1}
\{\Phi^2_{1},  \Phi^2_{\alpha_1}  \}&  = & 0 \\
\{\Phi^2_{\alpha_1},  \Phi^2_{\beta_1} \}&  = &
f^2_{ {\alpha_1} {\beta_1} 1}  \Phi^2_{1}
+
f^2_{ \alpha_1 \beta_1 \gamma_1}  \Phi^2_{\gamma_1} \label{eq:alg12}
\eea
The structure functions \(f^2_{ \alpha_1 \beta_1 \gamma} \) of algebra
(\ref{eq:alg1}),(\ref{eq:alg12}) possess the significant property :
 due to the fact
that  \( \Phi^2_1 \) is the  central element  the structure functions obey
the following property:
\bea \label{eq:integr1}
\{\Phi^2_{1},  f^2_{\alpha_1 \beta_1 \gamma}  \}&  = & 0
\eea
To verify this, it is enough to calculate the Poisson bracket
of \(\Phi^2_{1} \) with (\ref{eq:alg12}) and use the Jacobi identity .

 To extend the center of the new algebra \( {\cal A}^2 \) by \( \Phi^2_2\),
we will act by analogy with the previous case
\begin{itemize}
\item[a)]  exclude  \( \Phi^2_1\) and  \( \Phi^2_2\)
>from the left hand side of eq. (\ref{eq:alg12})
\item[b)] then achieve the  abelianization with all others
\end{itemize}

To carry out the first point of this program we will deal with two
consecutive transformations \( {\cal S}^3 \) and \({\cal S}^4\).
Let us require that the first transformation  \( {\cal S}^3 \)  of type
(\ref{eq:s})
\bea
\Phi^3_1 & = & \Phi^2_1 \nn\\
\Phi^3_2 & = & \Phi^2_2 \nn\\
\Phi^2_{\alpha_2} &  = & \Phi^2_{\alpha_2} + C^3_{\alpha_2}\Phi^2_{1},
\quad
\alpha_2 = 3,\dots, m
\eea
lead to the new algebra of constraints so that
\(\Phi^3_1 \) is again the central element
\bea
\{\Phi^3_{1},  \Phi^2_{\alpha_1}  \}&  = & 0
\eea
and
the  Poisson brackets of the second constraint \( \Phi^3_2 \) with
all other modified constraints does not contain  \( \Phi^3_1 \)
\bea
\{\Phi^3_{2},  \Phi^2_{\alpha_2} \}&  = &
\sum_{\gamma \not= 1} f^3_{ 2 \alpha_2 \gamma}  \Phi^3_\gamma .
\eea
This requirement leads to the   following equations
\bea \label{eq:2}
\{ \Phi^{3}_{1}, C_{\alpha_2}^{3} \} &=& 0 \nn\\
\{ \Phi_2^{2}, C_{\alpha_2}^{3} \}& = &
f^{2}_{2 \alpha_2 \gamma_2}
  C_{\gamma_2}^{3}  - f^{2}_{2 \alpha_2 1}
\eea
What about the existence of the solutions to these equations.
One can verify  that the integrability condition for the
system of differential equations (\ref{eq:2})  is nothing else but
(\ref{eq:integr1}).
 In full analogy with the previous case one can
express the new structure functions \(f^3_{\alpha, \beta \gamma} \)
through \(f^2_{\alpha, \beta \gamma} \)  and verify
that they obey the following property
\bea \label{eq:integr2}
\{\Phi^3_{1},  f^3_{ 2 \alpha_2 \gamma}  \}&  = & 0
\eea
Now one can realize the transformation  \( {\cal S}^4 \)  of type
(\ref{eq:s})
\bea
\Phi^4_1 &= & \Phi^3_1 \nn\\
\Phi^4_2 &= & \Phi^3_2 \nn\\
\Phi^4_{\alpha_2}& = &\Phi^3_{\alpha_2} + C^4_{\alpha_2}\Phi^2_{2}
\eea
so that  \(\Phi^4_1 \) is again central  element
\bea
\{\Phi^4_{1},  \Phi^2_{\alpha_1}  \}&  = & 0
\eea
and
the  Poisson brackets of the second constraint \( \Phi_4^2 \) with
all other modified constraints do not contain  \( \Phi^4_1 \)
and \( \Phi^4_2 \)
\bea
\{\Phi^4_{2},  \Phi^4_{\alpha_2} \}&  = &
\sum_{\gamma \not= 1,2} f_{ 2 \alpha_2 \gamma}  \Phi^3_\gamma .
\eea
This requirement leads to the  following equations :
\bea \label{eq:3}
\{ \Phi^{4}_{1}, C_{\alpha_2}^{4} \} &=& 0 \nn\\
\{ \Phi^3_{2}, C_{\alpha_2}^{4} \}& = &
f^{3}_{2 \alpha_2 \gamma_2}
  C_{\gamma_2}^{4}  - f^{3}_{2 \alpha_2 2}
\eea
This system is consistent in virtue of equations (\ref{eq:integr2}) .
For the new structure function one can again to verify that
 \bea \label{eq:integr3}
\{\Phi^4_{1},  f^4_{ 2 \alpha_2 \gamma}  \}&  = & 0
\eea
as for the previous step (see eq.(\ref{eq:integr2}).
Now for abelianization of two constraints \(\Phi_1^4, \Phi_2^5 \)
it is enough to perform last transformation with matrix
\({\cal R} \) of type  (\ref{eq:r}) if its  elements are
the solution to  the equations
\bea \label{eq:4}
\{\Phi^4_{1},
B^5_{\alpha_2\beta_2} \}& = & 0\nn\\
\{\Phi^4_{2},  B^5_{\alpha_2\beta_2} \}& = & - f^4_{2 {\gamma_2}
\beta_2}B^5_{\alpha_2\gamma_2}
\eea
As a result
\[ \{\Phi^5_{1},
\Phi^5_{\alpha} \} =  \{\Phi^5_{2}, \Phi^5_{\alpha} \} = 0
\]
Note that (\ref{eq:integr3})
provides the existence of the solution to eq.(\ref{eq:4}).
As a result, one can easy verify that the new structure
functions possess the property
\bea \label{eq:integr4}
\{\Phi^5_{{\overline{\alpha}}_2},  f^5_{ \alpha_2\beta_2 \gamma}  \}
=  0 ,
\quad  {\overline{\alpha}}_2 = 1, 2.
\eea
Thus, in five steps
(for summary, see Table 1.).
we  obtain an equivalent to initial
algebra \({\cal A}^5 \) with two central elements
\(
\Phi^5_{{\overline{\alpha}}_2} \)

Now let us suppose that by acting in such
a way we get the algebra \({\cal A}^{k-1} \)
(see Table 2.)
\bea \label{eq:algk}
\{\Phi^{a_k-1}_{\alpha},  \Phi^{a_k-1}_{\beta}  \}&  = &
f^{a_k -1}_{\alpha \beta \gamma} \Phi^{a_k-1}_{\gamma}
\eea
with the center  composed by \(k-1\) elements
\[
 {\cal Z}_{k-1} = \left( \Phi_1^{a_k-1}, \Phi_2^{a_k-1}, \dots
\Phi_{k-1}^{a_k-1} \right)
\]
\[
\{ {\cal Z}_{k-1}, \Phi_\alpha^{a_k-1} \} = 0
\]
and the structure functions of this algebra
have the  property
\bea \label{eq:intc}
\{\Phi^{a_k-1}_{\overline{\alpha}_{k}},  f^{a_k-1}_{\alpha_k\beta_k
\gamma}  \}&  = & 0 , \quad   {\overline{\alpha}_{k}}= k-1, \dots, m
\eea
Now by direct calculations it is easy to verify that  via the
action of transformation
of the matrix \({\cal D}^k\) with elements which are the solutions to eqs.
(\ref{eq:lin1}) and  (\ref{eq:lin2}),   we obtain the algebra
\( {\cal A}^{k}\)  with the center  composed by \(k\) elements
\bea {\cal Z}_{k-1} =
\left( \Phi_1^{a_k-1}, \Phi_2^{a_k-1}, \dots, \Phi_{k-1}^{a_k-1} \right) \nn\\
\{ {\cal Z}_{k-1}, \Phi_\alpha^{a_k-1} \} = 0.
\eea
The conditions
(\ref{eq:intc}) play the role of the integrability conditions for the
system of eqs.(\ref{eq:lin1}), (\ref{eq:lin2}).

For completion we must only to prove
the following property of structure functions
\bea \label{eq:intck}
\{\Phi^{a_k+k}_{\overline{\alpha}_{k+1}},  f^{a_k+k}_{\alpha_{k+1} \beta_{k+1}
\gamma}  \}&  = & 0
\eea
To verify  this one can consider the algebra of constraints
\bea \label{eq:alg}
\{\Phi^{a_k+k}_{\alpha_{k+1}},  \Phi^{a_k+k}_{\beta_{k+1}}  \}&  = &
f^{a_k +k}_{ \alpha_{k+1} \beta_{k+1} {\overline{\gamma}_{k+1}}}
\Phi^{a_k+k}_{{\overline{\gamma}}_{k+1}}
+
f^{a_k +k}_{ \alpha_{k+1} \beta_{k+1} {\gamma}_{k+1}}
\Phi^{a_k+k}_{{\gamma}_{k+1}}
\eea
and compose the Poisson bracket of (\ref{eq:alg})
with \(\Phi^{a_k+k}_{{\overline{\alpha}}_{k+1}} \).
Taking  into account that
\( \Phi^{a_k+k}_{{\overline{\alpha}}_{k+1}} \) are the central elements of
the algebra \({\cal A}^k\),  we immediately
get the desired result (\ref{eq:intck}) with  the help of Jacobi identity.

\section{\sc Christ and Lee model}
\setcounter{equation}0

In this section we will apply  the above described procedure to
the well known example --  nonabelian    Christ and Lee model
described by  the Lagrangian
\[
{\cal L} ({\bf x}, \dot{\bf x}, {\bf y}) =
\frac{1}{2}\left(\dot{\bf x}- [{\bf y},{\bf x}] \right)^2 - V({\bf x}^2)
\]
where \({\bf x}\) and \({\bf y}\) - are the three- dimensional vectors,
(\(x_1, x_2, x_3), (y_1, y_2, y_3\)).

It is easy to verify that except for  three primary constraints
\[
{\bfpi} = \frac{\partial{\cal L}}{\partial{\dot{\bf y}}} = 0,
\]
there are  two independent  constraints
\bea
\Phi^0_1 &= & x_2 p_3 -x_3 p_2 \nn\\
\Phi^0_2 & =& x_3 p_1 -x_1 p_3
\eea
with the algebra
\bea
\{\Phi^0_1 ,\Phi^0_2 \} &= &
- \frac{x_1}{x_3}\Phi^0_1 - \frac{x_2}{x_3}\Phi^0_2
\eea
The abelianization procedure for this simple case consists in  two stages
At first step, the transformation \({\cal S}^1\) reduces to
\bea
\Phi^1_1 & = & \Phi^0_1 \nn\\
\Phi^1_2  & = & \Phi^0_2  + C\Phi^0_1
\eea
and equation (\ref{eq:lin1}) looks like
\bea
\{\Phi^0_1 ,C \} & =&
 \frac{x_2}{x_3} C + \frac{x_1}{x_3}
\eea
One can write down a particular  solution to  this equation
\be
C(x) =  \frac{x_1}{x_3} \arctan{\left( \frac{x_2}{x_3} \right)}
\ee
So, as a result of first step we get a new algebra
\bea
\{\Phi^1_1 ,\Phi^1_2 \} &= &
- \frac{x_2}{x_3}\Phi^1_2
\eea
Now let us perform the  second transformation \( {\cal R}^2 \)
\bea
\Phi^2_1 & = & \Phi^1_1 \nn\\
\Phi^1_2  & = & B \Phi^1_2
\eea
with the function \(B \)  satisfing the equation of type (\ref{eq:lin2})
\bea
\{\Phi^1_1 ,B \} & =& \frac{x_2}{x_3}
\eea
A particular  solution to  this equation reads
\be
B(x) = \ln \left( \frac{\sqrt{x_2^2+x_3^2}}{x_3} \right )
\ee
Thus,  the equivalent to the initial abelian constraints have  the form
\bea
\Phi^2_1 & = & x_2 p_3 -x_3 p_2  \\
\Phi^2_2  & = & \ln \left(\frac{\sqrt{x_2^2+x_3^2}}{x_3} \right )
\left[ (x_3 p_1 -x_1 p_3) +
 \frac{x_1}{x_3} \arctan{\left( \frac{x_2}{x_3} \right)}
(x_2 p_3 -x_3 p_2)
\right]\nn
\eea

\section{\sc Concluding remarks}

We have discussed the iterative procedure of converting  first class
constraints in an arbitrary  singular  theory to the Abelian form of
constraints.
The final goal of dealing with a valuable form  of abelianization
is the construction
of the reduced phase space in the complicated  non - Abelian theory .
The application of our procedure to the  SU(2) Yang - Mills  will be
done in separate forthcoming publication.

\section{\sc Acknowledgments}

We would like to thank
 A.T. Filippov, G.T. Gabadadze, A.N. Kvinikhidze,
V.V. Nesterenko, G. Lavrelashvili for
 helpful and critical discussions.

This work was
supported in part by the Russian Foundation  of Fundamental Investigations,
Grant No 95\--02\--14411.

\newpage
\setcounter{page}{20}

\noindent Table 1. Abelianization stages for the algebra with
two central elements

\bigskip

\noindent\begin{minipage}[l]{250mm}
\textwidth 250mm
\textheight 230mm
\renewcommand{\arraystretch}{1.5}
\begin{tabular}{|l|l|l|l|} \hline\hline
&\multicolumn{1}{|c|}{
CONSTRAINTS} &  \multicolumn{1}{|c|}{ALGEBRA} &
\multicolumn{1}{|c|}{CONDITIONS}  \\\hline\hline
\hline
\hline
$\Phi_\alpha^{0}$
       &      $\Phi_\alpha^0 =\varphi_\alpha $
          &   $ \{ \Phi^0_\alpha,  \Phi^0_\beta \} =
                 f^0_{\alpha \beta \gamma} \Phi^0_\gamma $
               &   \\
\hline
\hline
$\Phi_\alpha^{1}$
       &   $ \Phi_1^1 = \Phi^0_1 $
        &   $\{ \Phi_1^1,  \Phi^1_{\alpha} \} =
               f^1_{1 \alpha \gamma_1} \Phi^1_{\gamma_1} $
            &   $\{ \Phi_1^0, C_{\alpha_1}^{1}  \} =  f_{1 \alpha_1 \gamma_1}
                 C_{\gamma_1}^{1}  - f_{1\alpha_1 1} $    \\
&  $\Phi^1_{\alpha_1} = \Phi^0_{\alpha_1} + C_{\alpha_1}^{1} \Phi^0_{1}$
            &
                &   \\
\hline
$\Phi_\alpha^{2}$
            &  $ \Phi_1^2 = \Phi_1^1 = \Phi^0_1 $
               & $\{\Phi_1^2,  \Phi^1_{\alpha} \} = 0 $
                 & $\{\Phi_1^1, B_{\alpha_1\beta_1}^{2} \} =
                - f^1_{1\gamma_1\beta}B_{\alpha_1\gamma_1}^{2} $    \\
&  $  \Phi_{\alpha_1}^{2} =
               B_{\alpha_1\beta_1}^{2}  \Phi^1_{\beta_1}$
                      &  &    \\
\hline
\hline
1 & \multicolumn{3}{|c|}{ $ \{ \Phi_{1}^{2}, \Phi_{\alpha}^{2} \} = 0 $ }\\
\hline
\hline
$\Phi_\alpha^{3}$
       & $ \Phi_1^3 = \Phi_1^2 = \Phi^1_1 $
         &$\{ \Phi_1^3,  \Phi^3_{\alpha} \} = 0$
              & $\{ \Phi_1^3, C_{\alpha_2}^{3} \} = 0 $   \\
       & $ \Phi_2^3 = \Phi_2^2  $
         & $\{ \Phi_2^3,  \Phi^3_{\alpha} \} =
           f^{3}_{1 \alpha \gamma_1} \Phi^2_{\gamma_1} $
              & $\{ \Phi_2^2, C_{\alpha_2}^{3} \} =  f^2_{2\alpha_2 \gamma_2}
                 C_{\gamma_2}^{2}  - f^2_{2\alpha_2 1} $    \\
&  $ \Phi_{\alpha_2}^{3} = \Phi_{\alpha_2}^2 + C_{\alpha_2}^{3}
    \Phi_{1}^1$
           &    &   \\
\hline
$\Phi_\alpha^{4}$
       & $ \Phi_1^4 = \Phi_1^3 ... = \Phi^1_1 $
           &$\{ \Phi_1^4,  \Phi^4_{\alpha} \} = 0 $
              & $\{ \Phi_1^4, C_{\alpha_2}^{4}  \} = 0 $    \\
       & $ \Phi_2^4 = \Phi_2^3 = \Phi_2^2  $
        & $\{ \Phi_2^4,  \Phi^4_{\alpha} \} =
           f^{4}_{2\alpha \gamma_2} \Phi^4_{\gamma_2} $
              & $\{ \Phi_2^3, C_{\alpha_2}^{4}  \} =  f^{3}_{2 \alpha_2
\gamma_2}
                 C_{\gamma_2}^{4}  - f^3_{2\alpha_22} $   \\
&  $ \Phi_{\alpha_2}^{4} = \Phi_{\alpha_2}^3 + C_{\alpha_2}^{4}
    \Phi_{2}^3$
           &    &   \\
\hline
$\Phi_\alpha^{5}$
            & $ \Phi_1^5 = \Phi_1^4 \cdots  \Phi_1^1 $
               & $\{ \Phi_1^5,  \Phi^5_{\alpha} \} = 0 $
                   &$ \{ \Phi_1^4, B_{\alpha_2\beta_2}^{5}  \} = 0 $ \\
            & $ \Phi_2^5 = \Phi_2^4 \cdots  \Phi_2^2 $
               & $\{ \Phi_2^5,  \Phi^5_{\alpha} \} = 0 $
                  & $\{ \Phi_2^4, B_{\alpha_2\beta_2}^{5}  \} =
            - f^4_{2 \gamma_2 \beta_2}B_{\alpha_2 \gamma_2}^{5} $    \\
          &  $  \Phi_{\alpha_2}^{5} = B_{\alpha_2\beta_2}^{5}
\Phi^4_{\beta_2}$
                      &  &    \\
\hline
\hline
2 & \multicolumn{3}{|c|}{ $ \{ \Phi_{1}^{5}, \Phi_{\alpha}^{5} \} =
 \{ \Phi_{2}^{5}, \Phi_{\alpha}^{5} \} =  0 $}\\
\hline
\hline
\end{tabular}
\end{minipage}

\newpage

\noindent Table 2. Abelianization stages for the algebra with
\(k\) central elements

\bigskip

\noindent\begin{minipage}[l]{230mm}
\oddsidemargin-35mm
\evensidemargin-35mm
\textheight 230mm
\renewcommand{\arraystretch}{1.5}
\begin{tabular}{|l|l|l|l|} \hline\hline
&\multicolumn{1}{|c|}{
CONSTRAINTS} &  \multicolumn{1}{|c|}{ALGEBRA} &
\multicolumn{1}{|c|}{CONDITIONS}  \\\hline\hline
\hline
\hline
k-1 & \multicolumn{3}{|c|}{ $ \{ \Phi_{1}^{a_k-1}, \Phi_{\alpha}^{a_k-1} \} =
 \{ \Phi_{2}^{a_k-1}, \Phi_{\alpha}^{a_k-1} \} =  \dots =\{ \Phi_{k-1}^{a_k-1},
\Phi_{\alpha}^{a_k-1} \} = 0 $}\\
\hline
\hline

       & $ \Phi_1^{a_k} = \Phi_1^{{a_k}-1} \cdots = \Phi^1_1 $
        & $\{ \Phi_1^{a_k}, \Phi^{a_k}_{\alpha} \} = 0 $
	 & $\{ \Phi_1^{a_k}, C^{a_k}_{\alpha_k} \} = 0$  \\
       & $ \Phi_2^{a_k} = \Phi_2^{a_k-1}\dots =\Phi_2^2  $
         & $\{ \Phi_2^{a_k}, \Phi^{a_k}_{\alpha} \} = 0$
                 & $\{ \Phi_2^{a_k}, C^{a_k}_{\alpha_k} \} = 0$ \\
        & \vdots   & & \\
       & $ \Phi_k^{a_k} = \Phi_k^{a_k-1} $
         & $\{ \Phi_k^{a_k}, \Phi^{a_k}_{\alpha} \} =
            f^{a_k}_{k \alpha \gamma_1} \Phi_{\gamma_1}^{a_k} $
	 & $\{ \Phi_k^{a_k-1}, C^{a_k}_{\alpha_k} \} =
                 f^{a_k-1}_{k \alpha_k \gamma_k}
                 C_{\gamma_k}^{{a_k}}  - f^{a_k-1}_{k\alpha_k 1} $    \\

&  $ \Phi_{\alpha_k}^{a_k} = \Phi_\alpha^{a_k-1} + C_{\alpha_k}^{a_k}
    \Phi_{1}^1$
           &    &   \\
\hline
&\vdots &\vdots&\vdots\\
&\vdots &\vdots&\vdots\\
\hline

       & $ \Phi_1^{a_k+k} = \cdots =\Phi_1^1 $
               & $\{ \Phi_1^{a_k+k},  \Phi_{\alpha}^{a_k+k} \} = 0 $
                 & $\{ \Phi_1^{a_k+k}, B_{\alpha_k\beta_k}^{a_k+k} \} = 0$
\\
        & $ \Phi_2^{a_k+k} = \cdots =\Phi_2^2 $
               & $\{ \Phi_2^{a_k+k},  \Phi_{\alpha}^{a_k+k} \} = 0 $

 & \\
&\vdots &\vdots&\vdots\\
      & $ \Phi_k^{a_k+k} =  \Phi_k^{a_k-1} $
         & $\{\Phi_k^{a_k+k},\Phi^{a_k+k}_{\alpha}\}= 0 $
         & $\{B_{\alpha_k\beta_k}^{a_k+k},\Phi_k^{a_k+k}\}=
f^{a_k+k-1}_{k\gamma_k\beta_k}B_{\alpha_k\gamma_k}^{a_k+k} $ \\
&  $  \Phi_{\alpha_k}^{a_k} =
               B_{\alpha_k\beta_k}^{a_k+k}  \Phi^{a_k+k-1}_{\beta_k}$
                      &  &    \\
\hline
\hline
k & \multicolumn{3}{|c|}{ $ \{ \Phi_{1}^{a_{k+1}-1}, \Phi_{\alpha}^{a_{k+1}-1}
\} =
 \{ \Phi_{2}^{a_{k+1}-1}, \Phi_{\alpha}^{a_{k+1}-1} \} =  \dots =
\{ \Phi_{k}^{a_{k+1}-1},
\Phi_{\alpha}^{a_{k+1}-1}\} = 0 $}\\
\hline
\hline
\end{tabular}
\end{minipage}

\end{document}